\documentclass[a4paper]{article}

\usepackage[utf8]{inputenc}
\usepackage[T1]{fontenc}

\usepackage[a4paper,top=2cm,bottom=2cm,left=2cm,right=2cm,marginparwidth=2cm]{geometry}
\usepackage{amsmath,amssymb}
\usepackage{graphicx}
\usepackage[colorinlistoftodos]{todonotes}
\usepackage{hyperref}
\hypersetup{colorlinks,allcolors=black}
\usepackage{authblk}
\usepackage{soul}
\usepackage{enumitem}
\usepackage[numbers]{natbib}
\usepackage{titlesec}
\usepackage{subfigure}
\usepackage{float}
\usepackage{tabularx}
\usepackage{booktabs}
\titlespacing\section{0pt}{12pt plus 4pt minus 2pt}{0pt plus 2pt minus 2pt}
\titlespacing\subsection{0pt}{12pt plus 4pt minus 2pt}{0pt plus 2pt minus 2pt}
\titlespacing\subsubsection{0pt}{12pt plus 4pt minus 2pt}{0pt plus 2pt minus 2pt}

\usepackage{blindtext}

\title{Optimal reaction coordinates and kinetic rates from the projected dynamics of transition paths}

\author[1]{Line Mouaffac}
\author[1]{Karen Palacio-Rodriguez}
\author[1]{Fabio Pietrucci\footnote{fabio.petrucci@sorbonne-universite.fr}}
\affil[1]{Sorbonne Université, Musée National d'Histoire Naturelle, UMR CNRS 7590, Institut de
Minéralogie, de Physique des Materiaux et de Cosmochimie, IMPMC, F-75005 Paris,
France}

\date{}                     

\begin{document}
\maketitle




\begin{abstract}
Finding optimal reaction coordinates and predicting accurate kinetic rates for activated processes are two of the foremost challenges of molecular simulations.
We introduce an algorithm that tackles the two problems at once:
starting from a limited number of reactive molecular dynamics trajectories (transition paths), we automatically generate with a Monte Carlo approach a sequence of different reaction coordinates that progressively reduce the kinetic rate of their projected effective dynamics. Based on a variational principle, the minimal rate accurately approximates the exact one, and it corresponds to the optimal reaction coordinate.
After benchmarking the method on an analytic double-well system, we apply it to complex atomistic systems: the interaction of carbon nanoparticles of different sizes in water. 
\end{abstract}

\newpage
\section{Introduction}

Physico-chemical transformations such as phase transitions, chemical reactions, and biomolecular conformational changes are characterized by metastable states separated by free-energy barriers, so that transitions between states are rare events.
Atomistic computer simulations of rare events -- especially molecular dynamics (MD) -- can play an important role, complementary to experiments, by predicting mechanisms, free-energy landscapes, and kinetic rates. 

However, MD simulations face two prominent challenges. First,
a gap -- often of many orders of magnitude --
between the short time scale of atomic motion (femtoseconds) and the long time scale of rare events hampers direct, brute-force simulations.
Second, the high-dimensional nature of configuration space makes the analysis of the atomistic trajectories and the extraction of relevant information intrinsically difficult.~\cite{pietrucci2017,glielmo2021} 

To overcome or at least alleviate these challenges and gain insights into the transformation processes, including quantitative thermodynamic and kinetic properties, it is necessary to reduce the dimensionality of the problem by projecting on an appropriate low-dimensional space. To this aim, collective variables (CVs) -- generic functions of atomic coordinates able to track interesting structural changes -- are often introduced, heuristically or by machine learning.~\cite{Gkeka20}

Among all possible CVs describing a transition, the optimal CV, or ``reaction coordinate" (RC), is widely considered the so-called committor function: 
in MD simulations, this function is defined as the probability $p_B(x)$ that a system will reach state $B$ before state $A$ starting from atomic positions $x$ in $3N$-dimensional configuration space, with initial momenta randomly drawn from the equilibrium distribution~\cite{bolhuis2002transition,rhee2005one,Metzner06,E10,banushkina16,peters16,zhang2016effective}.

The committor is considered optimal since i) it allows predicting the fate of atomic configurations towards reactants or products, and ii) it preserves the kinetics when employed to build a reduced model of the dynamics ({\sl vide infra}). 
The concept dates back to the 1930's, when  Onsager studied the recombination of a pair of ions in the presence of a uniform electric field \cite{Onsager38}. 

The committor can be used to design tests for the ``quality" of a RC: for instance, a good RC $q$ is expected to display a distribution of values $p_B(x)$ sharply peaked at 0.5 when considering a set of different configurations $x$ corresponding to a same location $q(x)=q*$ in RC space, i.e., precisely identifying the transition state configurations.~\cite{bolhuis2002transition,Hummer04,best2005reaction} 



Some remarks are in order. Given a generic CV $q$, the corresponding free energy landscape is always well-defined mathematically, via the equilibrium density $F(q)=-kT\log \rho_{\mathrm eq}(q)$: if the CV is able to resolve (at least partially) reactants and products, such $F(q)$ will display a barrier. 
However, different CVs can contain different amounts of information about the transition mechanism. 
Moreover, the height of the barrier will depend on the particular choice of the CV~\cite{jungblut2016pathways,dietschreit2022free} (see Figure~\ref{fig:dw-cvs-opt}). 

The barrier along the optimal RC is supposed to contain more useful information from the viewpoint of the calculation of the rate, keeping in mind that any invertible transformation of the optimal RC results again in an optimal RC, since such transformation cannot produce or destroy microscopic information: in the case of a non-linear transformation, the free-energy barrier and the diffusion coefficient are different for the two RCs~\cite{Krivov08,jungblut2016pathways}, the differences compensating each other in such a way that the projected dynamics keeps the same kinetic properties.

It is therefore important to distinguish the problem of converging the calculation of the free energy barrier along any given CV from a statistical viewpoint (a sampling problem) from the problem of identifying among all CVs the optimal RC and computing the corresponding barrier, for the purpose of estimating the rate.

Several methods, based on different theoretical principles, have been put forward to estimate optimal RCs from MD data. 
Given the customary identification of the optimal RC with the committor function, a possibility consists in optimizing a parametrization of the committor function based on $p_B(x)$ values estimated from transition path sampling (TPS), exploiting, e.g., the histogram test \cite{best2005reaction}, maximum-likelihood approaches \cite{peters07,peters16}, or artificial intelligence \cite{ma2005automatic,jung19}. 

We remark, however, that precisely estimating committor values very close to zero or to one by shooting requires an unfeasible large amount of trajectories: this hampers the accurate reconstruction of the optimal RC in presence of large free-energy barriers elsewhere than in the vicinity of the barrier top. 
This problem is addressed in the reweighted path ensemble approach~\cite{rogal10}. 

Several approaches have been developed to generate flexible RC representations able to capture slow degrees of freedom connected to rare events~\cite{perez2013,tiwary2016spectral,noe2017collective,m2017tica,perez2018advances,wu2020variational,he2022committor}.
Starting from the seminal work of Ma and Dinner~\cite{ma2005automatic}, where a neural network was trained with committor data,
recent years saw a flourishing of machine-learning approaches to RC optimization.~\cite{ribeiro2018reweighted,chen2018molecular,sidky2020,hooft2021,sun22}

In this work we address with a single methodology several prominent tasks of molecular simulations: identifying an optimal RC, estimating the free energy landscape and the diffusion along such RC, and estimating the rate of the process.

Results in the literature indicate that i) a model of the projected dynamics along the optimal RC gives more direct access to accurate kinetic rates \cite{best2005reaction,banushkina16,peters16}, and ii) the accurate rate is the minimal one -- attained for the optimal RC -- with respect to the rates computed from the projected dynamics along each of all possible CVs \cite{zhang2016effective}.
The method we present exploits these principles: putative CVs are generated in a Monte Carlo scheme, and the kinetic rates of associated effective low-dimensional models are estimated and systematically minimized, leading to optimal RCs and accurate kinetic predictions. 
An important ingredient of our approach is the automatic construction of stochastic (Langevin) models of CV evolution: for this purpose, we adopt the maximum likelihood algorithm proposed in Ref. \citenum{palacio2022free}, based on short TPS-like MD trajectories. We remark that such task is non-trivial from a numerical viewpoint \cite{Ferretti20}, and alternative approaches exist \cite{hummer05,zhang11,boninsegna2018,sicard21}.







\section{Theoretical methods}

\subsection{Variational principle: the optimal coordinate provides the minimal rate}

We start by considering ergodic diffusive dynamics (as described by overdamped Langevin equations) in a high-dimensional space, for a system with two metastable states $A$ and $B$. 
The transition rate from state $A$ to $B$ is denoted by $k_{A \rightarrow B}$. 
Next, we consider the effective dynamics resulting from the projection of the full high-dimensional dynamics onto a low-dimensional manifold of collective variables (CVs): 
such effective dynamics is an approximation of the original one, formulated again in terms of diffusive equations~\cite{legoll10,zhang2016effective}.
the reaction rate of the effective dynamics between states $\Tilde{A}$ and $\Tilde{B}$, defined in the low-dimensional space, is $\Tilde{k}_{\Tilde{A} \rightarrow  \Tilde{B}}$. Zhang, \textit{et al.}\cite{zhang2016effective} proved that the transition rate of the full dynamics is always smaller or equal to the one computed using effective dynamics. In other words, the optimal reaction coordinate yields a minimal rate,

\begin{equation}\label{eq:k-var-pple}
    k_{A \rightarrow B} \leq \Tilde{k}_{\Tilde{A} \rightarrow \Tilde{B}}
\end{equation}

The optimal projection, preserving the value of the rate constant
$k_{A \rightarrow B}= \Tilde{k}_{\Tilde{A} \rightarrow \Tilde{B}}$,
is achieved when the CV corresponds to the committor function or any invertible function of the latter.

As a complementary result, Ref.~\cite{Lu14} formally proved that it is possible to obtain accurate rates from an overdamped Langevin model when  projecting high-dimensional underdamped Langevin dynamics on a single CV, the committor, irrespective of the existence of timescale separation and metastability in the system. Considering that typical MD simulations, where Hamilton's equations are coupled with a thermostat, are akin to high-dimensional underdamped Langevin equations, the previous result should hold also in conventional MD.

Our approach consists in optimizing a RC defined as a function of several trial CVs by minimizing the kinetic rate of the projected dynamics with respect to all possible definitions of the RC. 
Here we consider the projection on a one-dimensional CV, assuming that the variational principle connecting the optimal CV to the minimal rate holds  also when the input data is represented by MD simulations with a thermostat. 
The projected dynamics is approximated by an overdamped Langevin equation, consistently with Refs.~\cite{zhang2016effective,Lu14}:
\begin{equation}\label{eq:OLEposdep}
\dot q = -\beta D(q) \frac{\partial F(q)}{\partial q} +\frac{\partial D(q)}{\partial q} +\sqrt{2 D(q)}\,\eta(t)~,
\end{equation}
with $D(q)$ the diffusion coefficient, $F(q)$ the free energy profile and $\eta(t)$ a Gaussian white noise of zero mean and unit variance. 
The use of this stochastic model has several appealing features: the only parameters of the equation consist in the functions $F(q)$ and $D(q)$, the mass and velocities are not explicitly needed, the Markovian character allows for simple likelihood expressions, and the mean first passage time (MFPT) of the $A \rightarrow B$ transition, inverse of the kinetic rate $k$, can be directly obtained by numerical integration \cite{Zwanzig01,jungblut2016pathways}: 
\begin{equation}\label{eq:tau}
    k^{-1}= \int_{q_0}^{b} dx\, \frac{e^{\beta F(x)}}{D(x)} \int_{a}^{x} dy\, e^{-\beta F(y)}
\end{equation}
where $q_0$ is the starting point in $A$ for the MFPT computation, while $a$ and $b$ are respectively the reflecting and absorbing boundaries. 

\subsection{Algorithm for reaction coordinate optimization}

The phase-space information used for RC optimization is represented by MD trajectories spanning transitions between the reactants and products regions of configuration space.
We adopt TPS as a source of input data for several reasons:~\cite{Hummer04,best2005reaction,peters16} i) ergodic trajectories (spontaneously spanning the transitions) are unfeasible in presence of free-energy barriers $\gg k_BT$, while TPS has an affordable cost for many systems; ii) TPS trajectories are free from biasing forces and, at statistical convergence, faithfully reproduce ergodic trajectories; iii) recent numerical evidence indicate that $\sim 100$ TPS-like MD trajectories projected on a suitable CV are sufficient to reconstruct accurate free-energies and rates by means of overdamped Langevin models~\cite{palacio2022free}.

The proposed RC optimization algorithm starts from the projection of TPS trajectories on a basis set of $N$ potentially-relevant CVs ${\bf Q}(t)=(Q_1(t),Q_2(t)...Q_N(t))$.
All these coordinates are put on the same footing by shifting and scaling so that for each of them the range of variation on the actual MD trajectories is between 0 and 1 (oriented so that $A\rightarrow B$ for growing values).
Note that such a linear transformation of the coordinate does not deform the corresponding free energy landscape besides an irrelevant additive constant:
passing from $x$ to $y=ax+b$ the probability densities transform according to $e^{-\beta [F(x)-\tilde{F}(y)]}=\rho(x)/\tilde{\rho}(y)=dy/dx=a$. Moreover, the new diffusion coefficient is scaled,  $\tilde{D}_y/D_x=(dy/dx)^2=a^2$~\cite{risken1996fokker}, as can be easily seen in the simple case of a driftless constant-$D$ diffusion process: $\langle [x(t)-x(0)]^2 \rangle=2D_x t$ implies that $\langle [y(t)-y(0)]^2 \rangle=2(a^2 D_x)t=2\tilde{D}_y t$.


A trial RC $q=
\sum_1^N w_i\,Q_i\equiv 
{\bf w}\cdot{\bf Q}$  
as a linear combination of the basis-set CVs is generated from random weights
normalized as ${\bf w}^2=1$. 
A Monte Carlo loop is then started, at each step proposing a new RC $q_{\text{new}}$
obtained by modifying the weights of the last accepted step
through small random variations 
${\bf w}\rightarrow{\bf w}+\delta{\bf w}$ (each $\delta w_i$ being drawn from a uniform distribution between $[-0.05,0.05]$). 

A newly proposed RC is accepted or rejected based on a Metropolis criterion aimed at minimizing the kinetic rate  estimated from a maximum-likelihood Langevin model.
The latter is optimized following the method in Ref.~\cite{palacio2022free}:
for a sufficiently small time interval $\tau$, the transition probability density $p$ (propagator) between points $q$ and $q'$ in CV-space can be approximated as~\cite{Drozdov97}
\begin{equation}\label{eq:prop}
p(q',t+\tau | q,t) \approx \frac{1}{\sqrt{2\pi \mu}}
e^{-(q'-q-\phi)^2/2\mu}
\end{equation}
\begin{equation}\label{eq:prop2}
\phi = a\tau +\frac{1}{2}(a a' + D a'')\tau^2
\ \ ,\ \mu = 2D\tau +
(aD' + 2a'D + DD'') \tau^2
\end{equation}
where the prime indicates $d/dq$, $a=-\beta D F' + D'$ is the drift in Eq.~\ref{eq:OLEposdep}, and the approximation includes terms up to order $\tau^2$.
We verified that resorting to the less accurate first-order propagator $\phi = a\tau = (-\beta D F' + D')\tau$, $\mu = 2D\tau$ gives unsatisfactory results for the systems considered in this work.

Based on Eq.~\ref{eq:prop2}, the likelihood of the MD trajectory, sampled with a time resolution $\tau$ and projected on a CV $q$ 
is given by
\begin{equation}\label{eq:logL}
-\log \mathcal{L}(\theta)=
\sum_{k=1}^{M-1} \Bigl\{ \frac{1}{2}\log[2\pi \mu_k(\tau)] 
+\frac{[q_{k+1}-q_k-\phi_k(\tau)]^2}{2\mu_k(\tau)} \Bigr\}
\end{equation}
where $\theta$ represents the set of all parameters of the Langevin equation (i.e., the profiles $F(q)$ and $D(q)$), $M$ is the number of trajectory configurations, and $\phi_k(\tau)\equiv\phi(q_k,\tau)$, $\mu_k(\tau)\equiv\mu(q_k,\tau)$.
The optimal Langevin model for the given $q$ and $\tau$, obtained by minimizing $-\log \mathcal{L}(\theta)$ as a function of the parameters with an iterative stochastic algorithm (see Ref.~\cite{palacio2022free} for details), directly yields the free energy and diffusion profiles.

The kinetic rate is computed from the Langevin model using Eq. \ref{eq:tau},
with integral boundaries defined in the following way: $a$ is the smallest observed value of $q$ (for each CV the transition $A\rightarrow B$ is in the direction of increasing $q$ values), while $q_0$ and $b$ are the average final values of $q$ for shooting trajectories ending in $A$ and $B$, respectively, i.e., the bottom of the corresponding free-energy minima.

Having obtained the rate corresponding to a given proposed CV, a Metropolis-like test is employed to accept or reject the CV based on the following expression for the probability: 
\begin{equation}\label{eq:probMC}
         P = \min\left[
         1\,,\,  \Big(\frac{k_{\text{old}}}{k_{\text{new}}}\Big)^\alpha  
         \times
    \frac{
    \tau^{\text{noise}}_{\text{old}}
    }{
    \tau^{\text{noise}}_{\text{new}}
    }
         \right]
\end{equation}
this expression tends to minimize the rate
(with $\alpha$ playing the role of an adjustable inverse Monte Carlo temperature, as discussed in the Results)
and simultaneously --
coherently with the hypothesis of a Markovian Langevin-like behavior --  the memory of the stochastic process. The latter is quantified by
$\tau^{\text{noise}}$, the auto-correlation time of the ``observed" noise trajectory $G_k$ estimated from the $q_k$ trajectory via the optimal Langevin model: 
\begin{equation}\label{eq:effnoise-em}
  G_k = \frac{q_{k+1}-q_k + 
  [ \beta D(q_k) F'(q_k) -D'(q_k) ] \tau}
  {\sqrt{2 D(q_k) \tau}}
\end{equation}
(see Ref.~\cite{palacio2022free} for details). 
 
To summarize, the algorithm performs the following steps:

\begin{enumerate}[noitemsep]
    \item Project TPS trajectories relaxing from a barrier top  onto a basis set of CVs $\{Q_i\}_{i=1,...,N}$ 
    \item Construct a first trial RC $q$ as a sum of the basis CVs with random weights
    
    \item Generate a new CV $q_{\text{new}}$ by randomly modifying the weights
    
    \item Estimate the free energy and diffusion profiles as well as the $A\rightarrow B$ rate from an optimal Langevin model of the  observed trajectory $q_{\text{new}}(t)$ using likelihood maximization 
    
    \item  Accept or reject the new CV with the probability in Eq.~\ref{eq:probMC}, aimed at minimizing the rate and enforcing Markovianity 
    
    \item  Go to step 3 (iterate until convergence of the rate)
\end{enumerate}

At convergence, the algorithm provides an optimal RC, as well as the corresponding free-energy and diffusion profiles and kinetic rate, all starting solely from a set of $~100$ TPS-like short MD trajectories. 
If the basis set contains all the relevant degrees of freedom for the transition process, the optimal rate should approach the exact rate. 

\subsection{Analytic double-well potential}

A two-dimensional double-well free-energy surface is defined, for benchmark purposes, as the sum of two Gaussian-shaped wells: 
\begin{equation}\label{eq:dw-cvs-opt}
F(x,y)=-C\left[
e^{-\frac{ (x-x_0)^2 }{ 2\sigma_x^2 }}
e^{-\frac{ (y-y_0)^2 }{ 2\sigma_y^2 } } +
e^{-\frac{ (x-x_1)^2 }{ 2\sigma_x^2 }}
e^{-\frac{ (y-y_1)^2 }{ 2\sigma_y^2 } }
\right]    
\end{equation}
with $C=20\,k_BT$, Gaussian centers
$(x_0,y_0)$=$(0.1723,0.5058)$, $(x_1,y_1)$=$(0.8060,0.5058)$, and widths 
$\sigma_x^2=0.03921$, $\sigma_y^2=0.2519$.

We consider the region $x \in [0,1]$ and $y \in [0,1]$, see Fig. \ref{fig:dw-cvs-opt}a.
The diffusion matrix was set to $\mathbf{D}=
0.015\cdot \big(\begin{smallmatrix}
  1 & 0\\
  0 & 1
\end{smallmatrix}\big)$~ps$^{-1}$. 

The reference MFPT was estimated directly from brute-force trajectories. For this purpose, 10 long overdamped Langevin simulations with a time step of 0.1~fs were performed. For the aggregate simulation time (10 $\mathrm{\mu}$s), 16~039 jumps from state $A$ to $B$ were observed, yielding a MFPT of $5600\pm 40$~ps and a transition rate $k_{A \rightarrow B} = 1.79\pm0.01 \cdot 10^{-4}$~ps$^{-1}$.

Input trajectories for RC optimization were obtained by shooting 100 overdamped Langevin trajectories from the barrier top. From these trajectories, 51 relaxed to state $A$ (left well) and 49 relaxed to state $B$ (right well). The basis set CVs, in this case, are simply $x$ and $y$, yielding two one-dimensional projections $x(t)$ and $y(t)$ of the 100 relaxing trajectories.

\subsection{Fullerene dimers in water}

The association and dissociation of C$_{60}$ and C$_{240}$ fullerene dimers (OPLS-AA force-field \cite{Jorgensen96}) in water solution (SPC force-field \cite{berendsen81}) have been simulated by MD using GROMACS v2019.4 \cite{Berendsen95, Abraham15} patched with PLUMED 2.5.3 \cite{Tribello14}.
The simulations are similar to those in Ref.~\citenum{palacio2022free}: for more computational details than what is summarized here, we refer to the latter article.

In the first system, two C$_{60}$ molecules were solvated by 2398 water molecules in a simulation box of $3.607^3$~nm$^3$ with periodic boundary conditions. 
In the second system, two C$_{240}$ molecules were solvated by 5375 water molecules in a simulation box of
$5.22^3$~nm$^3$.
MD simulations were performed with a time step of 1 fs in the $NPT$ ensemble at 298 K~\cite{Bussi07} (thermostat time constant = 1 ps) and 1 atm~\cite{Parrinello81} (barostat time constant = 4 ps).

The reference free-energy profiles of the association/dissociation of the C$_{60}$ fullerenes in water as a function of the 8 basis CVs 
were computed from unbiased simulations of 1~$\mathrm{\mu}$s total aggregated time, from the population 
histogram: $F(Q_i)=- k_BT \log\rho_\mathrm{eq}(Q_i)$. 
Error bars were estimated as the standard deviation of the mean over 5 independent replicas.

We used as input data for RC optimization a set of 100 aimless shooting~\cite{Peters06,peters07} trajectories of 20 ps, generated using the script publicly available at \url{https://github.com/physix-repo/aimless-shooting}, 
employing a separation of 0.1~ps between successive shooting points.

For the C$_{60}$ dimer we define
the dissociated state based on the distance between centers of mass as $d \geq 1.34$~nm and the associated state as $d \leq 1.17$~nm;
for the C$_{240}$ dimer we define
the dissociated and associated states as $d \geq 2.01$~nm and $d \leq 1.9$~nm, respectively.
The reference rate constants for the dissociation of the C$_{60}$ and C$_{240}$ dimers from the associated state were estimated using the reactive flux formalism over 1000 aimless shooting trajectories (see Ref. \citenum{palacio2022free} for details),
obtaining for C$_{60}$ a MFPT of $6.1 \pm 1.2$~ns 
and for C$_{240}$ a MFPT of $9.4 \pm 1.1$~$\mathrm{\mu}$s.

The basis set CVs employed for projecting the fullerene dimers trajectories are the following: 

\begin{enumerate}
    \item $d$: the distance between the centers of mass of each fullerene molecule.
    
    \item $cc$: the number of carbon-carbon contacts, estimated summing  continuous coordination functions between any atom of the first fullerene (set $S_1$) and any atom of the second fullerene (set $S_2$)  
    \begin{equation}\label{eq:ful-cvs-c}
    cc = \sum_{i \in S_1} \sum_{j \in S_2} C_{ij}\ ,\ \ \ 
     C_{ij} = \frac{1-\Big(\frac{r_{ij}}{r_0}\Big)^n}{1-\Big(\frac{r_{ij}}{r_0}\Big)^m}
    \end{equation}
    where $r_{ij}$ is the distance between atoms $i$ and $j$, with parameters $r_0=0.35$~nm, $n=6$ and $m=10$. 
    
    \item $c2w$: the number of carbon-water contacts, defined similarly to $cc$ in Eq.~\ref{eq:ful-cvs-c}, in this case with $r_0=0.6$~nm, set $S_1$ including all carbon atoms of the two fullerene molecules, and set $S_2$ including all oxygen atoms of the water molecules.
    
    \item $c1w$: the water-carbon contacts for a single fullerene molecule, defined as $c2w$ except for the inclusion of a single fullerene in set $S_1$.
    
    \item $sc$: the approximate carbon pair entropy, estimated using the expression~\cite{piaggi2017enhancing}
    \begin{equation}
    sc=-2\pi\rho k_B \int_0^{r_{\mathrm{max}}} \left [ g(r) \ln g(r) - g(r) + 1 \right ] r^2 \mathrm{d}r~,
    \end{equation}
    where $\rho$ is the density, $r_{\mathrm{max}}$ is a cutoff set to 0.65~nm, and $g(r)$ is the pair distribution function of carbon atoms, estimated via Gaussian kernels as
    \begin{equation}
    g(r) = \frac{1}{4 \pi N \rho r^2} \sum_{i\neq j} \frac{1}{\sqrt{2 \pi \sigma^2}} e^{-(r-r_{ij})^2/(2\sigma^2)}~,
    \end{equation}
    where $N$ is the number of carbon atoms and $\sigma = 0.01$~nm.
    
    \item $sw$: the approximate water pair entropy, estimated with the same equations and parameters as $sc$ applied to oxygen atoms.
    
    \item $ucc$: the Van der Waals carbon-carbon potential energy, computed over all carbon pairs.
    
    \item $ucw$: the Van der Waals carbon-water potential energy, computed between all the carbon atoms and all the solvent atoms.
\end{enumerate}

In the Results section, for each putative RC in the optimization algorithm, the likelihood of a Langevin model was maximized employing $3 \cdot 10^{6}$ and $2 \cdot 10^{6}$ iteration steps for the $C_{60}$ and $C_{240}$ fullerenes, respectively, adopting a time resolution $\tau=1$~ps for the projected MD trajectory. The latter was tested as sufficient to yield time-decorrelated noise when using the dimer center-of-mass distance $d$ as CV.

\section{Results and discussion}

\subsection{Two-dimensional double well}

\begin{figure}[!ht]
  \centering
\includegraphics[width=0.9\textwidth]{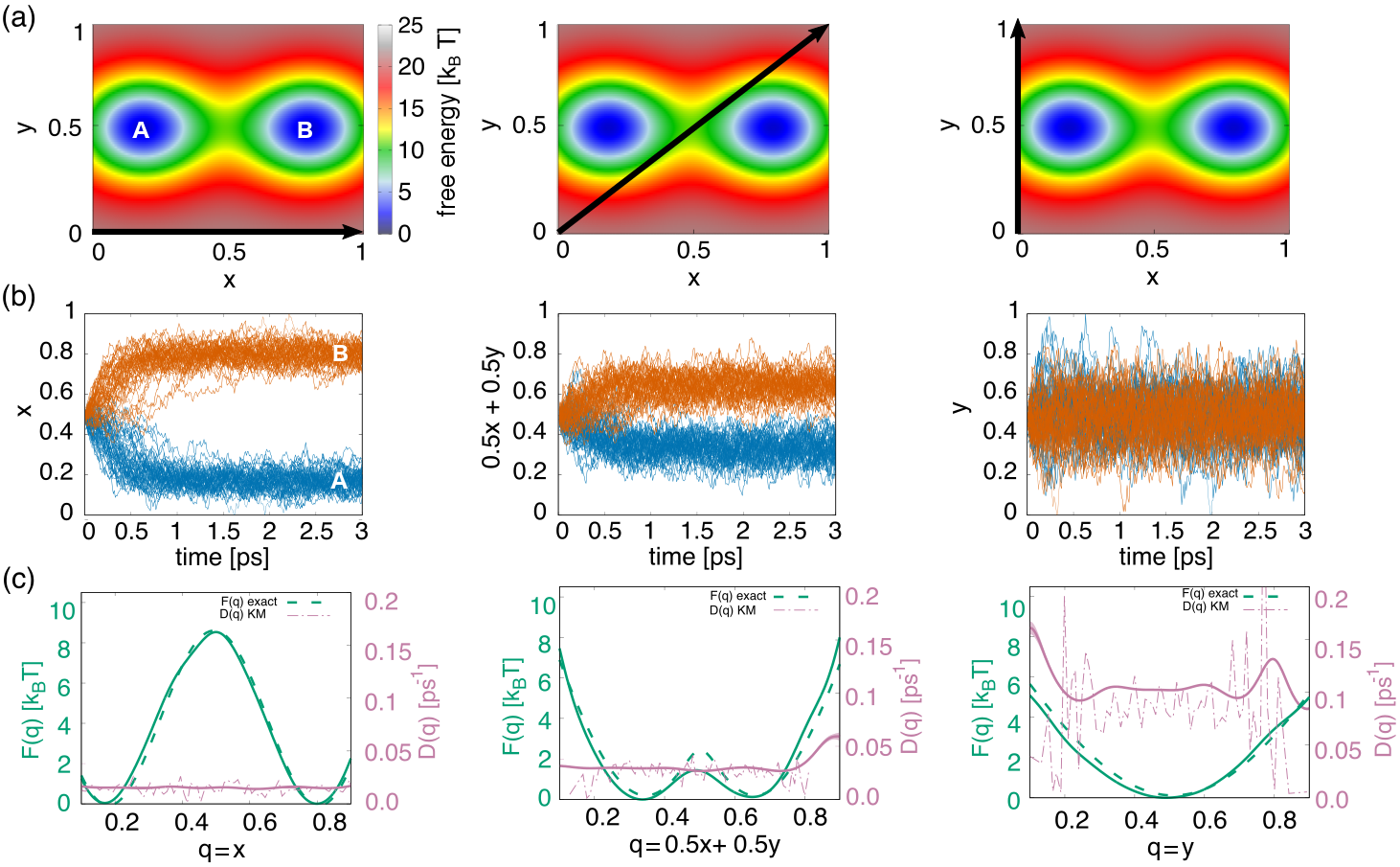}
  \caption{(a) Analytic double-well free-energy landscape, with arrows indicating one-dimensional projections along the $x$ axis, the diagonal, and the $y$ axis, respectively. (b) 100 Langevin bi-dimensional trajectories, projected on a single CV, relax from the barrier-top into $A$ (blue) or $B$ (orange), and are used to optimize effective one-dimensional Langevin models via likelihood maximization. (c) For each projection, the free-energy and diffusion profiles inferred from the latter models are shown (continuous lines with standard error over 10 reconstructions) alongside the exact $F(q)$ profiles and an alternative estimate of $D(q)$ (dashed lines), from the Kramers-Moyal formula~\cite{risken1996fokker}
  $D(q)\approx (\langle \Delta q^2 \rangle_q - \langle \Delta q \rangle_q ^2)/2\tau$
where $\Delta q$ is the CV displacement in a time interval $\tau$, with binned averages along $q$. 
  }
  \label{fig:dw-cvs-opt}
\end{figure}

To test the validity of the algorithm developed, we first benchmark it on an analytic model: the double-well system detailed in the Methods section. 
By construction the dynamics is of overdamped-Langevin nature, the trajectories being obtained by integration of such stochastic differential equation over the two-dimensional $F(x,y)$ landscape in Figure~\ref{fig:dw-cvs-opt} with a constant isotropic diffusion matrix.

After projecting the trajectories on a one-dimensional CV,
several non-trivial effects can be anticipated from theory.
Figure~\ref{fig:dw-cvs-opt} illustrates for instance three simple choices of the CV, corresponding to the $x$ axis, the $y$ axis, or the 45$^\circ$ diagonal.

Langevin trajectories shown in Figure~\ref{fig:dw-cvs-opt}(b) clearly resolve two distinct end-states
when adopting $q\equiv x$, while the separation becomes less clear upon deviation of $q$ from the $x$ axis until vanishing for $q\equiv y$.

The exact one-dimensional $F(q)$ profile obtained by projecting on a generic CV $q(x,y)$ can be computed from probability marginalization as
\begin{equation}
    e^{-\beta F(q)}=\int dx \int dy\,
    e^{-\beta F(x,y)}\, \delta[q-q(x,y)]
\end{equation}
Figure~\ref{fig:dw-cvs-opt}(c) shows that CV changes result in significant differences in the one-dimensional free-energy profiles. In particular the barrier is maximal for $q\equiv x$, it is strongly reduced for the diagonal, and it vanishes for $q\equiv y$; analogously, the distance between the two minima is largest for $q\equiv x$ and it vanishes for $q\equiv y$.
These observations support the intuitive idea that $x$ is the optimal RC for this problem, fully capturing the progress of the transition mechanism. Moreover, the reduced barriers for alternative CVs -- simply the effect of overlapping contributions of the two wells --  suggest faster rates for the corresponding effective dynamics. 
The effective one-dimensional diffusion coefficient $D(q)$ (Figure~\ref{fig:dw-cvs-opt}(c)) is also affected by projection, increasing when $q$ deviates from $x$, however its effect on the rate is linear, less important than barrier variations, which have an exponential effect. 

\begin{figure}[!ht]
    \centering
\includegraphics[width=0.5\textwidth]{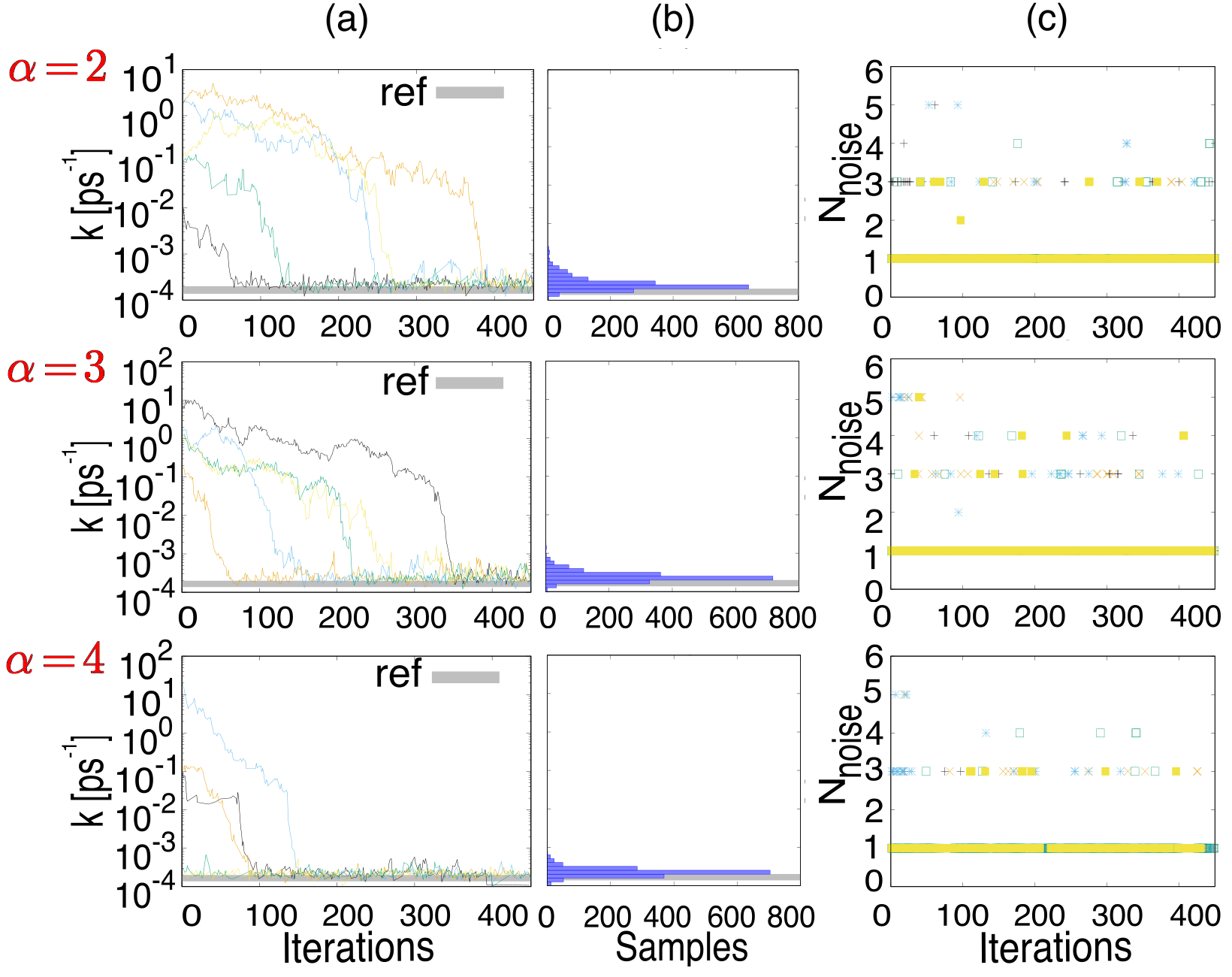}
    \caption{RC optimization by rate minimization applied to the double-well benchmark: (a) the rate evolution is shown for 5 independent optimizations with $\alpha=2,3,4$, respectively; (b) the histogram of the rate values in the stationary regime, i.e., discarding the initial relaxation; (c) number of correlated time steps (at O.O1~ps resolution) of the effective noise estimated from the original trajectories using the optimal Langevin model for each CV.}
    \label{fig:dw-results}
\end{figure}

Starting from $x$ and $y$ as basis CVs, 10 independent runs of the RC optimization algorithm are performed, aimed at minimizing the rate of the effective projected model (at time resolution $\tau=0.01$~ps), each starting with a different random linear combination as initial CV (see Figure~\ref{fig:dw-results}). 

The parameter $\alpha$ (effective Monte Carlo inverse temperature in Eq.~\ref{eq:probMC}) is varied from 2 to 4: 
in all tested cases the initial rate decreases almost monotonously, often by several orders of magnitude, until reaching an equilibrium behavior with small fluctuations in a small interval.
Low $\alpha$ values increase the acceptance of putative CVs in presence of an increase of the rate, whereas high $\alpha$ values decrease such acceptance causing the stochastic optimization to converge more rapidly towards a minimal rate, with smaller equilibrium fluctuations. 
$\alpha$ plays therefore the role of an adjustable parameter providing some control over the convergence of the rate minimization algorithm.

Irrespective of the initial putative CV the algorithm successfully and consistently retrieves an optimal combination including at least $95\%$ of $x$.
Given that after the initial relaxation a stationary distribution of equilibrated $k$ values is reached, with a finite width that depends on the parameter $\alpha$, introducing a degree of fuzziness, different criteria could be envisaged to identify the optimal rate values and the corresponding optimal RCs.
In the case of the double well system, identifying the predicted rate with  
the average value of the stationary distribution and its uncertainty with the standard deviation yields 
$2.6\pm 1.0\cdot 10^{-4}$~ps$^{-1}$ for $\alpha=3$, in good agreement with the reference brute-force rate of $1.8 \cdot 10^{-4}$~ps$^{-1}$.

However, the results for the solvated fullerene dimers (see next section), a realistic complex system, indicate the possibility of a few spurious outliers at the lowest $k$ values, resulting from imperfect optimization of the Langevin model for a few putative CVs. 
This suggests an alternative criterion to identify the optimal rate predicted by the algorithm, as the 5th percentile in the stationary distribution, i.e., the rate below which 5\% of the rates are found in the distribution.
This criterion predicts (for $\alpha=3$) a 5th percentile value of $1.6\cdot 10^{-4}$~ps$^{-1}$: taking as optimal prediction the average over the 5 rates closest to such value for each of the 10 independent optimization runs, we get a predicted rate of $1.60 \pm 0.04 \cdot 10^{-4}$~ps$^{-1}$
again in good agreement with the reference.

We remark that, even though the original dynamics in $(x,y)$-space is Markovian, 
non-Markovian effects could appear on the effective dynamics when projecting on CVs different from the committor, requiring in principle an underdamped (or generalized) Langevin description~\cite{Lu14}. 
Therefore, the term penalizing deviations from Markovianity in Eq.~\ref{eq:probMC} plays, in general, an important role, helping to reject proposed CVs that would introduce significant memory effects rendering inappropriate the overdamped model.
Moreover, as shown in Ref.~\citenum{palacio2022free}, from the viewpoint of the inference of Langevin models non-Markovian behavior appears to be correlated with barrier overestimation and thus rate underestimation, a spurious effect that must be avoided to exploit correctly the variational principle connecting the minimal rate to the optimal RC.

Our simulations show that for all values of $\alpha$ the vast majority of the CVs accepted during the optimization process have an effective noise free from time-correlation (Figure \ref{fig:dw-results}), with occasional exceptions, confirming the ability of the algorithm to enforce Markovianity.

\subsection{Interaction between fullerene nanoparticles in water}

\begin{figure}[!ht]
    \centering
    \includegraphics[width=0.5\textwidth]{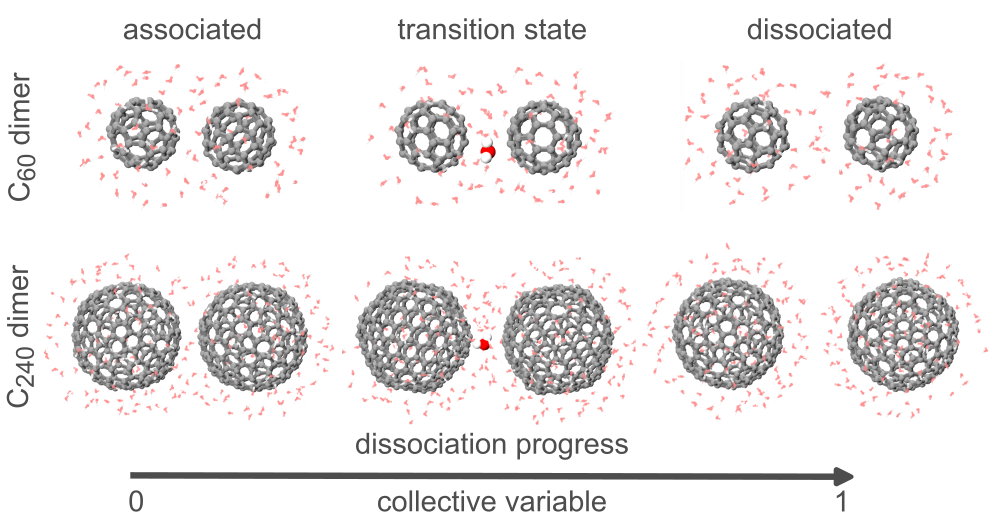}
    \caption{Characteristic atomic configurations for the C$_{60}$ and C$_{240}$ dimers in bulk water solution. Water molecules distant $<4$~\AA~from carbon atoms are shown. The progress from the associated state (left), passing through the transition state (center), until the dissociated state (right), is followed by CVs normalized to span between zero and one. A water molecule in the transition state, bridging between fullerenes and breaking the direct hydrophobic interactions between them, is highlighted.}
    \label{fig:dimers-rep}
\end{figure}

We applied the RC optimization algorithm to the case of the dissociation process of fullerene dimers in water solution, respectively C$_{60}$ and C$_{240}$. These systems are rather complex, including thousands of atoms, and feature an associated state, corresponding to a free-energy minimum, characterized by the carbon nanoparticles in close contact without the mediation of water molecules, and a dissociated state where each fullerene is fully-solvated. The transition state region features a water molecule bridging between the fullerenes (Figure~\ref{fig:dimers-rep}), corresponding to the top of a sizable free-energy barrier when a CV able to resolve the metastable states, like the distance $d$ between centers of mass, is employed~\cite{palacio2022free}.

We start by proposing a pool of eight basis CVs to be used as building blocks for RC optimization: the set ranges from geometry-inspired CVs like the distance $d$ and the total number of carbon-carbon or carbon-water contacts ($cc$, $c2w$, $c1w$) to physics-inspired CVs like the approximate two-body carbon or water entropy ($sc$, $sw$) and the carbon-carbon or carbon-water interaction energy ($ucc$, $ucw$). It is not obvious {\sl a priori} which CVs or combination thereof could be the best approximations of an optimal RC.

\begin{figure}[!ht]
    \centering
\includegraphics[width=0.45\textwidth]{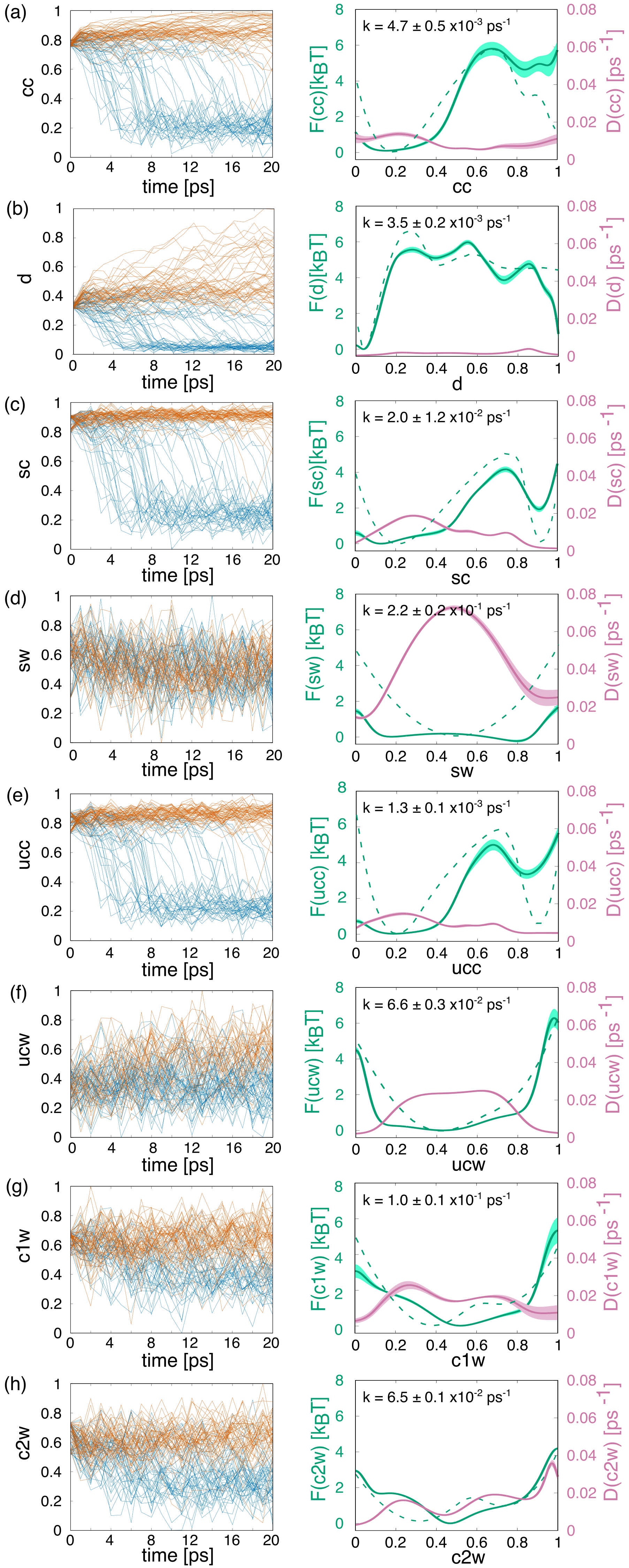}
    \caption{
    MD trajectories from TPS of the C$_{60}$ fullerene dimer in water relaxing to the associated (blue) or dissociated (orange) state (left), with the corresponding $F(q)$, $D(q)$ profiles and kinetic rates (right), for each CV of the basis set ((a) to (h), definitions in the Methods section).
    The profiles $F(q)$ and $D(q)$ with continuous lines are inferred from likelihood maximization (average and standard error over 10 independent runs). The kinetic rate $k$ is calculated using Eq.~\ref{eq:tau}. The dashed $F(q)$ are obtained from equilibrium MD histograms.
    }
    \label{fig:c60_profiles}
\end{figure}

The TPS trajectories relaxing from the barrier top, together with the free-energy and diffusion profiles inferred from likelihood maximization of Langevin models  for each of these CVs are reported for the C$_{60}$ dimer in Figure~\ref{fig:c60_profiles}, and display, as expected, strong variations depending on the CV choice (see the Supporting Information for C$_{240}$). 
Clearly, CVs able to resolve well the associated and dissociated states feature a barrier, contrary to CVs lacking such resolving power. Overall, the $F(q)$ landscapes compare well with the reference brute-force ones.
A time resolution $\tau=1$~ps is adopted for Langevin modelling, as it yielded Markovian behavior in combination with the CV $d$ for both fullerene sizes in Ref.~\citenum{palacio2022free}.

The basis CVs that results in the lowest kinetic rates, of the order of $10^{-3}$~ps$^{-1}$, are $d$, $cc$ and $ucc$; the CVs providing the highest (hence, less accurate) rates are instead $sw$ and $c1w$, of the order of $10^{-1}$~ps$^{-1}$. 
This allows a first ranking of the quality of the basis CVs.
The position-dependent diffusion coefficient $D(q)$ is roughly between $0.01$ and $0.03$~ps$^{-1}$ for all CVs except $sw$, that reaches almost $0.08$~ps$^{-1}$ (Figure~\ref{fig:c60_profiles}).

\begin{figure}[!ht]
  \centering
\includegraphics[width=0.99\textwidth,height=0.4\textwidth]{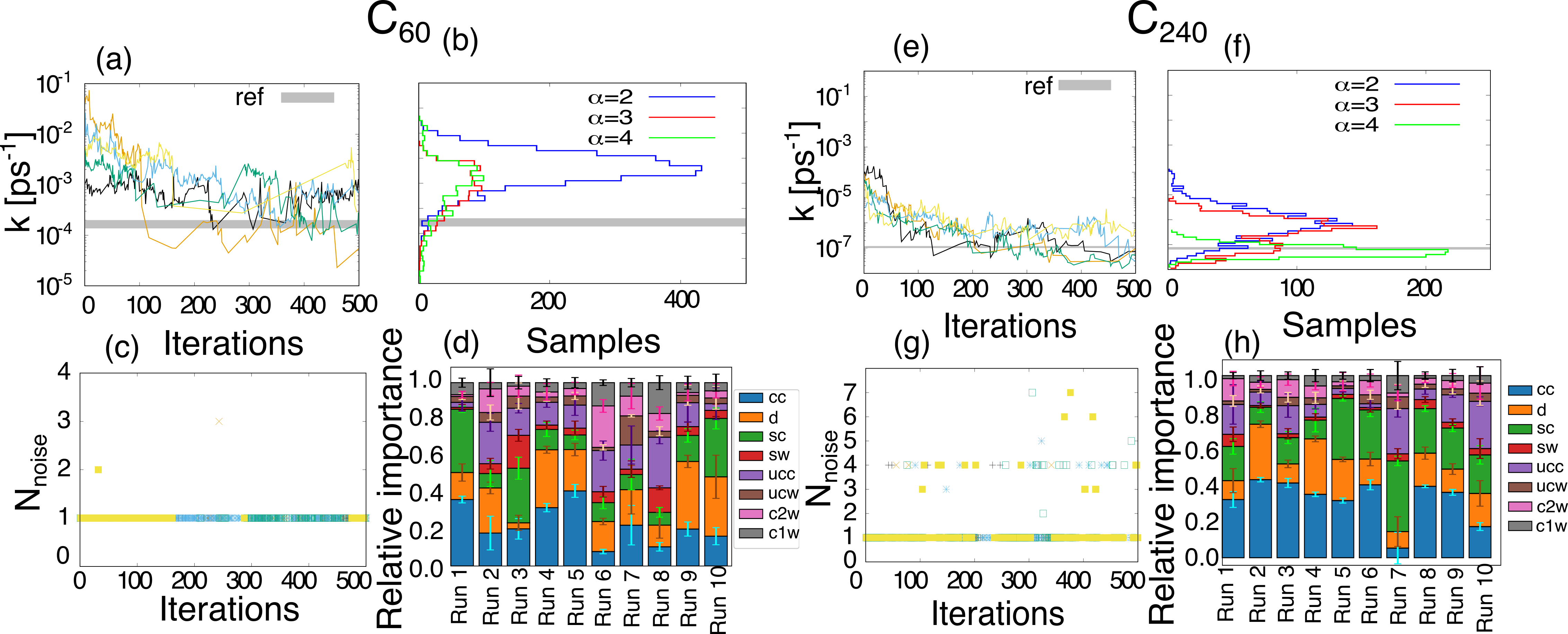}
  \caption{RC optimization results for the C$_{60}$ and C$_{240}$ fullerene dimers in water. (a,e) 5 independent rate minimization runs are displayed, the grey bar corresponding to the reference rate (thickness = error bar) estimated via reactive flux. (b,f) Histogram of the rate values from 10 optimizations in the stationary regime (i.e., discarding the first 300 iterations) for different $\alpha$ values. (c,g) Number of correlated time steps (at 1~ps resolution) of the effective noise. (d,h) Weights distribution for the optimal RCs from 10 independent runs:  for each run, the 5 RCs closest to the 5th percentile rate are included.}
  \label{fig:c60_opt}
\end{figure}

We applied the RC optimization algorithm (500 accepted steps, with $10^6$ Langevin optimization iterations for each putative CV) starting from different initial random combinations of the basis CVs and testing the parameter $\alpha=2,3,4$.
As shown in Figure~\ref{fig:c60_opt},
for both fullerene sizes the rate relaxes to a stationary distribution after about 200 steps: rate fluctuations are wider in these systems than for the double well, however the lower part of the distribution has the same order of magnitude of the reference rates.

Employing the 5th percentile of the stationary $k$ distribution as criterion (see previous section) at $\alpha=3$, we obtain as optimal rates the values $1.26\pm~0.15 \cdot 10^{-4}$~ps$^{-1}$ and $1.8\pm~1.1 \cdot 10^{-8}$~ps$^{-1}$  for the dissociation C$_{60}$ and C$_{240}$ dimers, respectively, that compare well with the references reactive-flux rates $1.6\pm~0.3 \cdot 10^{-4}$~ps$^{-1}$ and $1.1\pm ~0.1\cdot  10^{-7}$~ps$^{-1}$, respectively (see Methods section). 

The Markovianity appears enforced in an effective way, with the vast majority of the accepted CVs having uncorrelated effective noise and a few ones displaying $<10$ correlated steps (see Figure~\ref{fig:c60_opt}(c, g)).

Finally, the composition of the optimal RCs for the solvated C$_{60}$ and C$_{240}$ dimers is reported in Figure~\ref{fig:c60_opt}(d, h):
since the basis CVs are not uncorrelated between them, the optimal RC is not uniquely defined and different weights in the combination $q=
\sum_1^N w_i\,Q_i$ can lead to RCs with similar quality and similar rates. 

In fact, performing a principal component analysis on the members of the set of optimal RCs (i.e., the 50 RCs with rate closest to the 5th percentile in Figure~\ref{fig:c60_opt}(b,f)) provides explained variance ratios for the first four principal components $[0.40, 0.24, 0.14, 0.11]$ for C$_{60}$ and $[0.56, 0.17,0.16,0.04]$ for C$_{240}$ (see Supporting Information). 
This shows that it is not possible to easily reduce the dimensionality in the optimal space of RCs by means of one or two dominating components.

Nevertheless, some basis CVs contribute more than others, on average, to the optimal RCs: in particular, Figure~\ref{fig:c60_opt}(d,h) suggest that $d$, $cc$, $sc$, and $ucc$ play a prominent role for both fullerene sizes. The latter CVs do not contain water degrees of freedom, pointing, somehow counter-intuitively, to a limited role of the solvent in the optimal RC for these systems.

\section{Concluding remarks}

We presented an algorithm for the automatic optimization of the RC for activated processes starting from $\approx 100$ short TPS MD trajectories and a basis set of putative CVs.
The approach is based on the variational principle in Ref.~\citenum{zhang2016effective}: the rate predicted by a Langevin model of the effective projected dynamics is minimized by the optimal RC, coinciding in this limit with the true MD rate.

As an essential ingredient, we adopt the method in Ref.~\citenum{palacio2022free} to infer an optimal overdamped Langevin model for each putative RC, thus reliably recovering the corresponding free-energy and diffusion profiles as well as the kinetic rate. 

Since we make the assumption that an overdamped stochastic equation be able to faithfully model the projected dynamics, we explicitly disfavoured the appearance of memory effects in the acceptance probability expression applied to newly proposed CVs, thus keeping the Markovianity under control.

Numerical tests on an analytic double-well system as well as on MD simulations of carbon nanoparticles in explicit water solution indicate that optimal RCs are systematically identified in a robust and efficient way: this allowed us to predict kinetic rates in the microsecond time scale from $\sim 10$~ps-long trajectories.

A foremost advantage of the present algorithm is that RC optimization is done as post-processing of a MD data set: a large number of putative CVs can be screened with limited computer resources without the need to re-run expensive MD calculations. 
Moreover, compared to alternative approaches based on machine learning the committor, there is no need to estimate committor values over a large sample of atomic configurations, a computationally-expensive endeavour that, by construction, is limited to the region close to the barrier top.
On the contrary, the proposed approach only requires a small number of reactive trajectories and, by construction, it learns the optimal RC across the full span of the transition paths joining metastable states, even where the committor assume values very close to zero or to one.


Here the basis set of CVs is built from heuristic considerations, while in future works it could be advantageous to devise an unsupervised building procedure.
Likewise, from the simplistic formulation of optimal RCs as linear combinations of basis CVs, it would be beneficial for complex processes to adopt more flexible, nonlinear formulations, possibly including neural networks.
As a final word of caution, the effective description adopted here in terms of overdamped Langevin equations is appropriate for a subset of all possible rare-events systems and processes: underdamped or generalized (non-Markovian) stochastic models can be necessary in other cases.

We expect that the approach introduced here could help discovering optimal RCs and predicting accurate rates for challenging open problems like crystal nucleation and protein-protein or protein-ligand dissociation, whose complexity, so far, eluded systematic and predictive computational studies. 




\section{Acknowledgement}

We gratefully acknowledge very insightful discussions with Christoph Dellago, Gerhard Hummer, Hadrien Vroylandt, Arthur France-Lanord, Alessandro Barducci, Bettina Keller, Tony Lelièvre and Edina Rosta. 
This work was performed with the support of the Institut des Sciences du Calcul et des Données (ISCD) of Sorbonne University (IDEX SUPER 11-IDEX-0004). 
Calculations were performed on the GENCI-IDRIS French national supercomputing facility, under grant numbers A0110811069, A0120901387, A0130811069.




\clearpage

\section{Supporting information}


\begin{figure*}[htb!]
\centering
\includegraphics[width=0.5\textwidth]{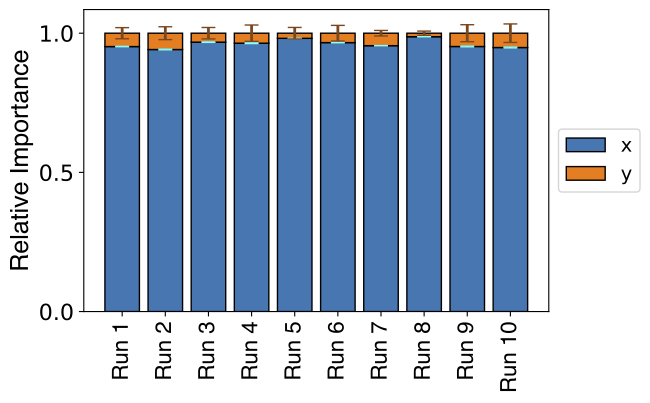}
\caption{Optimal RC composition for the benchmark double-well system with $\alpha=3$}
\end{figure*}

\begin{figure*}[htb!]
\centering
\includegraphics[width=0.45\textwidth]{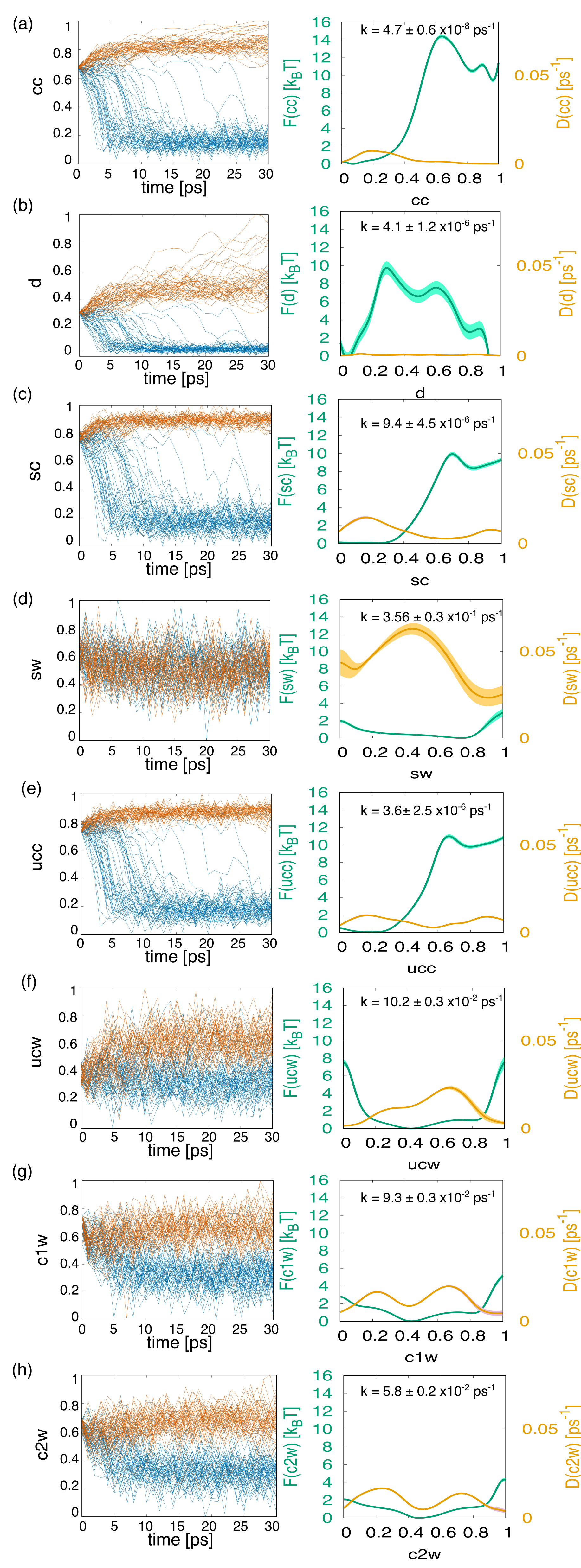}
\caption{
MD trajectories from TPS of the C$_{240}$ fullerene dimer in water relaxing to the associated (blue) or dissociated (orange) state (left), with the corresponding $F(q)$, $D(q)$ profiles and kinetic rates (right), for each CV of the basis set ((a) to (h), definitions in the Methods section). 
    The profiles $F(q)$ and $D(q)$ with continuous lines are inferred from likelihood maximization (average and standard error over 10 independent runs). The kinetic rate is calculated using Eq.~3.}
\end{figure*}

\begin{figure*}[htb!]
\centering
\includegraphics[width=0.6\textwidth]{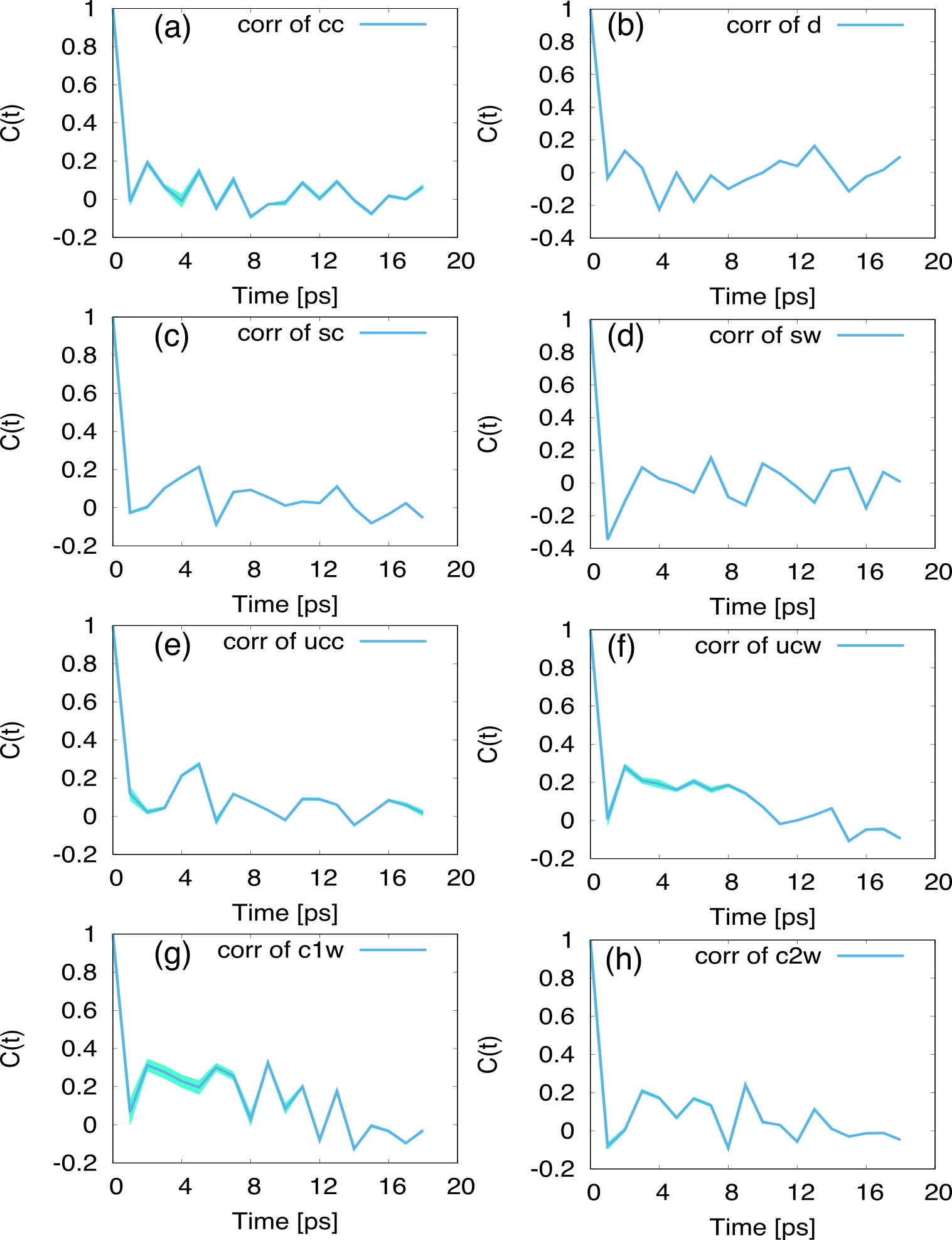}
\caption{Time auto-correlation of the effective noise of the CV basis set for C$_{60}$, evaluated on 100 TPS MD trajectories. The solid lines are the average correlation of 10 independent optimal models (see main text) while the shaded line represents the standard deviation.}
\end{figure*}

\begin{figure*}[htb!]
\centering
\includegraphics[width=0.6\textwidth]{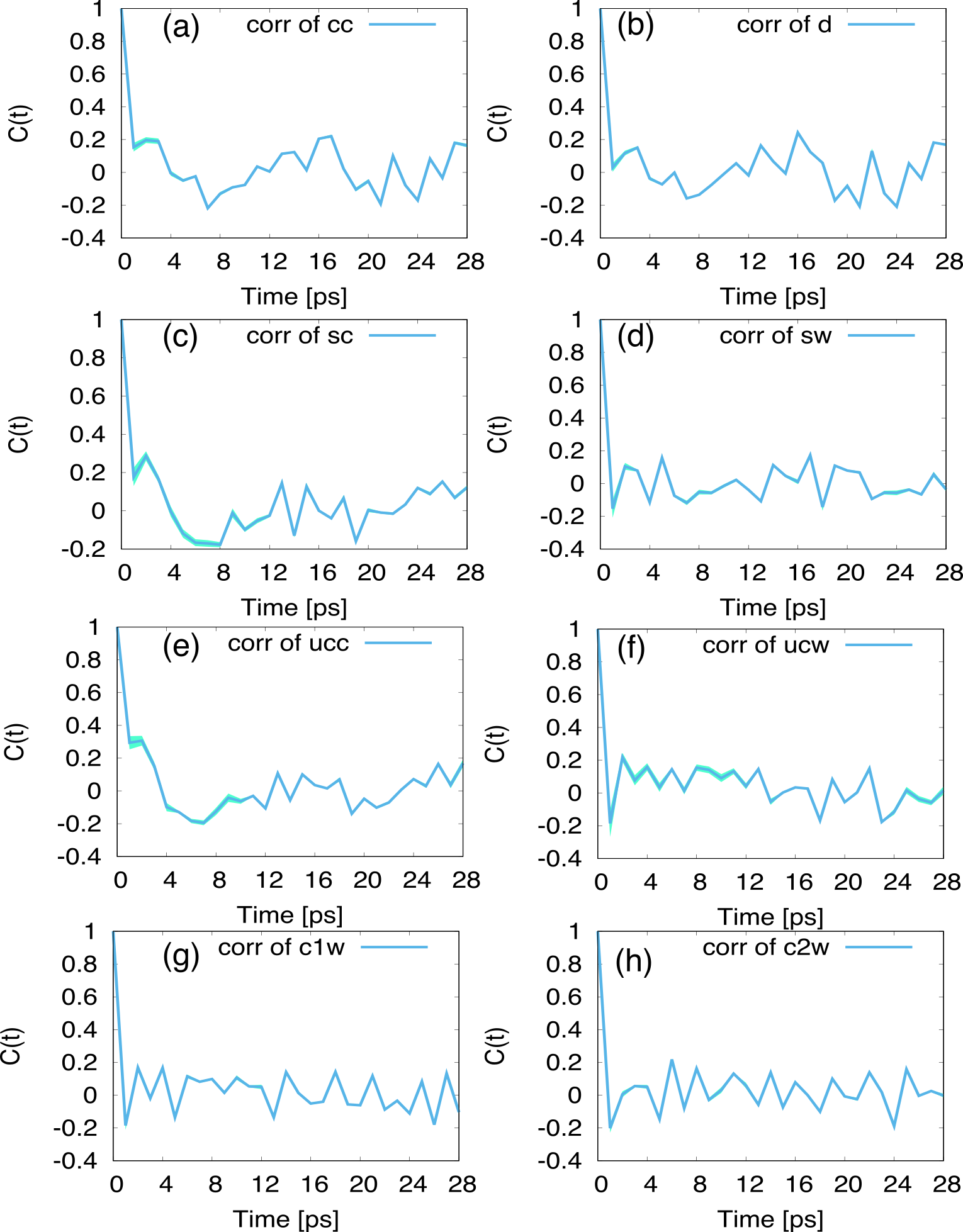}
\caption{Time auto-correlation of the effective noise of the CV basis set for C$_{240}$, evaluated on 100 TPS MD trajectories. The solid lines are the average correlation of 10 independent optimal models (see main text) while the shaded line represents the standard deviation.}
\end{figure*}


\begin{figure*}[htb!]
\centering
\includegraphics[width=0.7\textwidth]{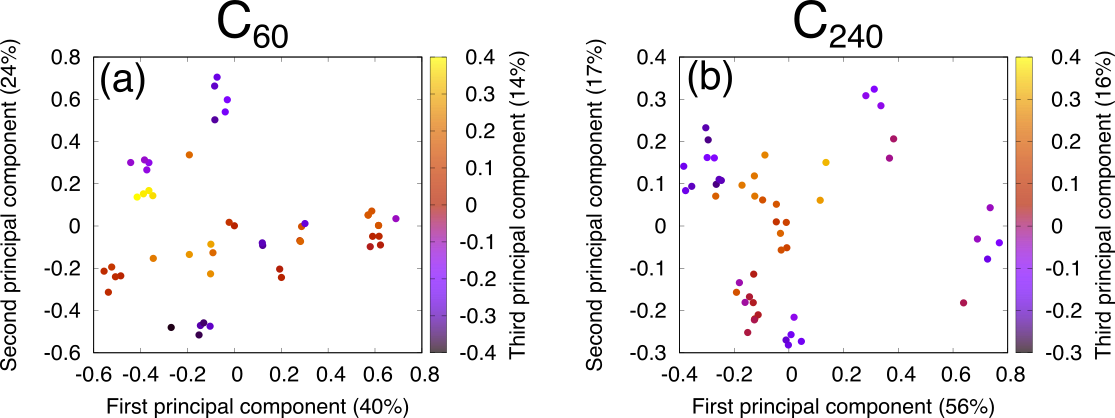}
\caption{Principal component analysis on the 50 members of the optimal set of RCs (a) for C$_{60}$ and (b) for C$_{240}$.}
\end{figure*}

\clearpage


\end{document}